\documentclass[aps, prl, preprint, amsmath,amssymb,showpacs,superscriptaddress]{revtex4-1}

\usepackage{graphicx}
\usepackage{color}
\usepackage{amsthm}
\usepackage{amsmath}
\usepackage{amssymb}
\usepackage{setspace}
\usepackage{braket}
\usepackage{float}
\renewcommand{\(}{\left(}
\renewcommand{\)}{\right)}

\makeatletter
\@ifundefined{textcolor}{}
{
 \definecolor{BLACK}{gray}{0}
 \definecolor{WHITE}{gray}{1}
 \definecolor{RED}{rgb}{1,0,0}
 \definecolor{GREEN}{rgb}{0,1,0}
 \definecolor{BLUE}{rgb}{0,0,1}
  \definecolor{purple}{rgb}{0.5,0,1}
 \definecolor{CYAN}{cmyk}{1,0,0,0}
 \definecolor{MAGENTA}{cmyk}{0,1,0,0}
 \definecolor{YELLOW}{cmyk}{0,0,1,0}
}

\newcommand{\E}{\mathcal{E}}

\newcommand{\bk}{\mathbf{k}}

\begin{document}

\title{Fermi arc mediated entropy transport in topological semimetals}

\author{Timothy M. McCormick}
\email[]{mccormick.288@osu.edu}
\affiliation{Department of Physics and Center for Emergent Materials, The Ohio State University, Columbus, OH 43210, USA}

\author{Sarah J. Watzman}
\affiliation{Department of Mechanical and Aerospace Engineering, The Ohio State University, Columbus, Ohio 43210, USA}

\author{Joseph P. Heremans}
\affiliation{Department of Physics and Center for Emergent Materials, The Ohio State University, Columbus, OH 43210, USA}
\affiliation{Department of Mechanical and Aerospace Engineering, The Ohio State University, Columbus, Ohio 43210, USA}
\affiliation{Department of Materials Science and Engineering, The Ohio State University, Columbus, Ohio 43210, USA}

\author{Nandini Trivedi}
\email[]{trivedi.15@osu.edu}
\affiliation{Department of Physics and Center for Emergent Materials, The Ohio State University, Columbus, OH 43210, USA}

\date{\today}

\begin{abstract}
In topological Weyl semimetals, the low energy excitations are comprised of linearly dispersing Weyl fermions, which act as monopoles of Berry curvature in momentum space and result in topologically protected Fermi arcs on the surfaces. We propose that these Fermi arcs in Weyl semimetals lead to an anisotropic magnetothermal conductivity, strongly dependent on externally applied magnetic field and resulting from entropy transport driven by circulating electronic currents.  The circulating currents result in no net charge transport, but they do result in a net entropy transport.  This translates into a magnetothermal conductivity that should be a unique experimental signature for the existence of the arcs.  We  analytically calculate the Fermi arc-mediated magnetothermal conductivity in the low-field semiclassical limit as well as in the high-field ultra-quantum limit, where only the chiral Landau levels are involved.  By numerically including the effects of higher Landau levels, we show how the two limits are linked at intermediate magnetic fields.  This work provides the first proposed signature of Fermi arc-mediated thermal transport and sets the stage for utilizing and manipulating the topological Fermi arcs in experimental thermal applications.
\end{abstract}
\pacs{}
\maketitle

\section{Introduction}

The discovery of topological band insulators has led to a new paradigm in condensed matter physics\cite{hasanKane,qiZhang}.  The recent theoretical prediction\cite{burkBal,wanTurnVish,PhysRevLett.107.186806,Volovik2014514} and subsequesnt experimental discovery\cite{Xu613,PhysRevX.5.031013,Lv2015} of topological Weyl semimetals has expanded the list of topologically nontrivial quadratic Hamiltonians.  These Weyl semimetals possess nodal Fermions composed of non-degenerate linear band crossings.  In order to satisfy the condition of nondegeneracy, Weyl semimetals must break either inversion-symmetry or time-reversal symmetry.  These so-called Weyl nodes come in pairs of opposite chirality\cite{Nielsen1981219} and, unlike the Dirac fermions in graphene\cite{graphene}, they are robust against the formation of a gap due to their three-dimensional nature.

The low energy linear dispersing modes of a Weyl semimetal carry monopole charges of Berry curvature.  There have been extensive predictions of novel electronic transport resulting from these Weyl nodes\cite{vishChargexport,sonSpivakWeyl,PhysRevB.88.125105,PhysRevLett.111.246603} as well as promising experimental signatures of some of these transport properties, such as negative longitudinal magnetoresistance \cite{PhysRevX.5.031023,Zhang2016,Arnold2016}.  This negative magnetoresistance in the presence of closely parallel electric and magnetic fields is a result of the local non-conservation of charge in the Brillouin zone as a consequence of the chiral anomaly in Weyl semimetals.

In addition to the plethora of novel transport phenomena exhibited by Weyl semimetals, these monopole charges of Berry curvature are responsible for topological Fermi arcs in Weyl semimetals, perhaps their most fascinating feature. Fermi arcs form open contours of surface states which terminate on the projections of Weyl nodes, or in the case of doped Weyl semimetals, terminate on the projections of Fermi pockets enclosing Weyl nodes.  Like the bulk Weyl nodes, the states comprising the Fermi arcs also disperse linearly.  These gapless boundary modes provide the key signature for Weyl semimetals in spectroscopy experiments.

Fermi arcs in Weyl semimetals are known to lead to exotic quantum oscillations involving mixed real and momentum space orbits\cite{pkv,Zhang2016a,Moll2016} as well as resonant transparency\cite{PhysRevX.5.041046}.
However, the effects of Fermi arcs on thermal transport so far remains a completely unexplored frontier.  Preliminary theoretical studies of thermal transport in Weyl semimetals have so far only considered contributions from the bulk Weyl fermions\cite{fieteThermoelec,PhysRevB.93.035116,doubleWeyl}.

In this Letter, we explore the role of Fermi arcs in entropy transport for the first time.
We demonstrate that Fermi arcs in topological semimetals lead to a thermal conductivity which depends strongly on an externally applied magnetic field.  This external magnetic field is shown to enhance the heat current in topological semimetals by creating "conveyor belt" motion of charge leading to a coherent transmission of heat.  
We propose that this Fermi arc-mediated entropic transport will lead to a dramatically anisotropic magnetothermal conductivity.
In the low-field semiclassical limit, it is shown that a Lorentz force on the electrons comprising the Fermi arcs necessarily leads to a bulk flow of electrons, carrying a net energy current in the presence of a temperature gradient.  When the field strength reaches the ultraquantum limit, we find that the electrons from the arcs can only hybridize with the chiral Landau levels in the bulk, which allow for a dissipationless flow of energy.  
In the intermediate quantum regime, where several quantized Landau levels become involved in Fermi arc-mediated magnetothermal transport, we calculate the specific heat and entropy density and the magnetothermal conductivity for various temperatures and magnetic fields.
We conclude by discussing the experimental implications of our work.  This work sets the stage for utilizing and manipulating the topological Fermi arcs in experimental magnetothermal applications.

\section{Fermi arc-mediated magnetothermal transport}

\begin{figure}
	\centering
	\includegraphics[width=0.5\textwidth]
	{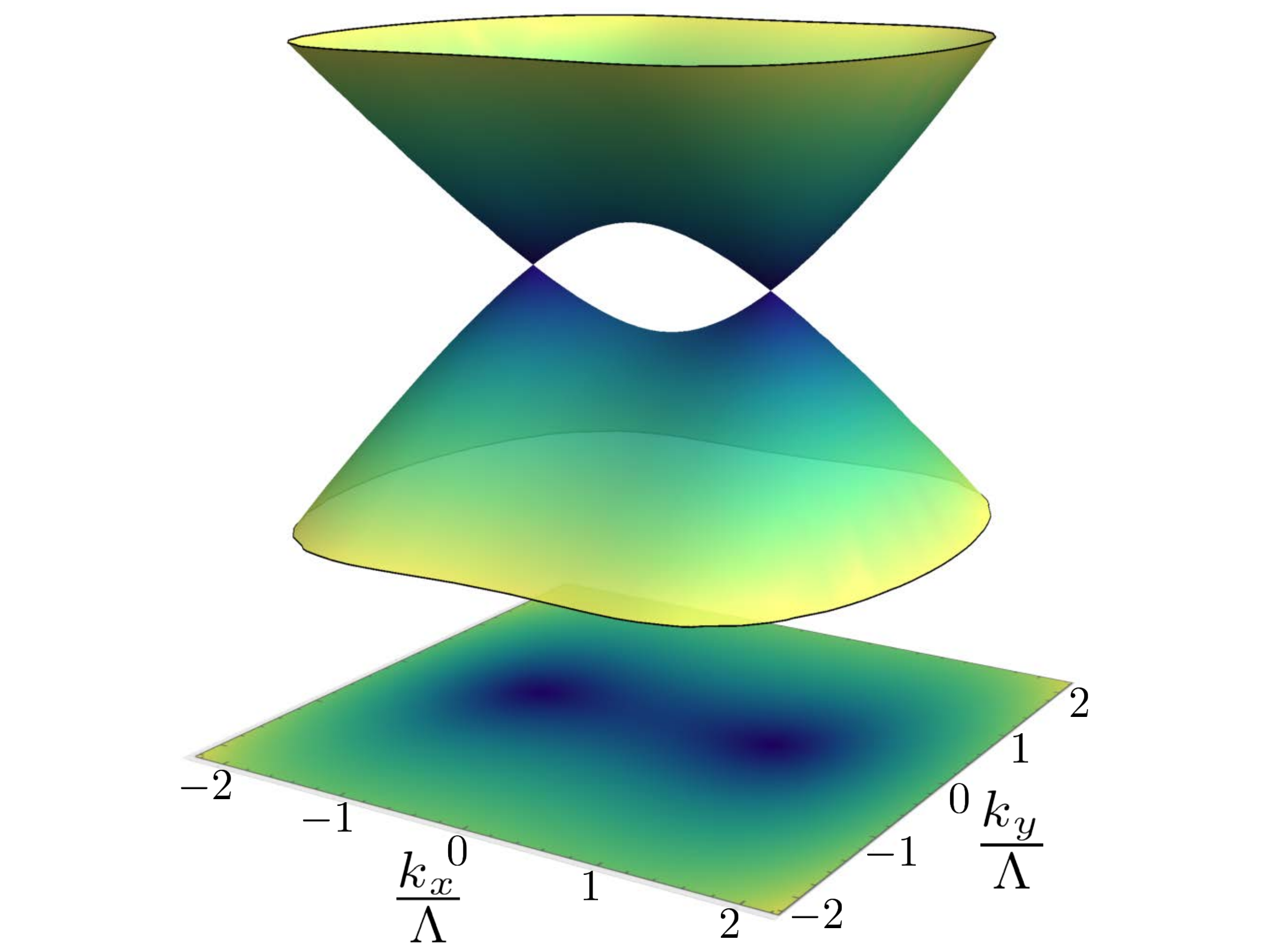}
	\caption{
	Energy dispersion of a pair of Weyl nodes given by Eqn. (\ref{contham}).  The Weyl nodes are separated by a distance of $2 \Lambda$ in the Brillouin zone.  Areas of the band structure colored in darker blue have a larger magnitude of Berry curvature.
	}
	\label{weylPlot}
\end{figure}

We consider a linearized model of Weyl fermions in the continuum limit.  For a given Weyl node of chirality $\chi = \pm 1$, the Hamiltonian that describes states near the node is given by 
\begin{equation}
\label{contham}
\hat{H}_\chi = \chi \hbar v_F 
\Big(
k_x \hat{\sigma}_x
+k_y \hat{\sigma}_y
+k_z \hat{\sigma}_z
\Big),
\end{equation}
where the Fermi velocity of the Weyl electrons is given by $v_F$ and where the Pauli matrices $\hat{\sigma}_j$ span a spin or orbital degree of freedom.  These Weyl nodes come in pairs of opposite chirality and we consider $N_p$ of such pairs of nodes.  We will take each pair of Weyl nodes to be separated in the Brillouin zone by a characteristic distance set by some momentum cutoff $\Lambda$.  We take all pairs of Weyl nodes to lie at the same energy $E = 0$, but our calculations generalize to cases where sets of Weyl nodes lie at different energies.

\subsection{Semiclassical limit}

We consider a Weyl semimetal in a slab geometry such that the Fermi arcs reside in the surface Brillouin zone labeled by $k_x$ and $k_y$ in the presence of a perpendicular magnetic field $\mathbf{B} = B\mathbf{e}_z$.  Then, assuming the Fermi arc states all have a uniform magnitude of velocity $v_F$, a semiclassical wavepacket of electrons on a Fermi arc will follow the trajectory
\begin{equation}
\label{arckdot}
\dfrac{d\mathbf{k}}{dt} = \dfrac{e}{\hbar c}\( \mathbf{v}_\bk \times \mathbf{B} \) = \dfrac{e}{\hbar c} v_F B \mathbf{e}_t,
\end{equation}
where $\mathbf{e}_t$ is the unit tangent vector to the Fermi arc.  This configuration has been predicted\cite{pkv} to give rise to unusual Fermi arc-mediated quantum oscillations in Weyl semimetals, but similar effects remain unexplored in thermal transport.

We consider the collective motion of the electrons comprising the Fermi arcs and note that the change in the number $N$ of electrons on a given surface with one Fermi arc is 
\begin{equation}
\label{dndtfa}
\dfrac{dN}{dt} = A \dfrac{dn}{d\E}\dfrac{d\E}{dk}\dfrac{dk}{dt},
\end{equation}
where $A$ is the area of the surface on which the Fermi arcs reside, $\frac{dn}{d\E}$ is the density of states of the Fermi arcs $g_{A}(\E)$, the magnitude of the Fermi velocity is of course $\frac{d\E}{dk} =\hbar v_F$ and, from Eqn. (\ref{arckdot}), we know that $\frac{dk}{dt} = \frac{e}{\hbar c} v_F B$.  The density of states on the Fermi arcs is given by
\begin{equation}
\label{gofefa}
g_{A}(\E) = \dfrac{k_0}{\hbar v_F},
\end{equation}
where $k_0$ is the length of the Fermi arc.  Here, we have made the reasonable assumption  that the magnitude of $v_F$ is constant along the arcs.
Hence we have that the total rate of electrons moving along each arc is given by 
\begin{equation}
\label{dndtfa2}
\dfrac{dN}{dt} = A \dfrac{e}{\hbar c}k_0 v_F B.
\end{equation}

\begin{figure*}
	\centering
	\includegraphics[width=0.99\textwidth]
	{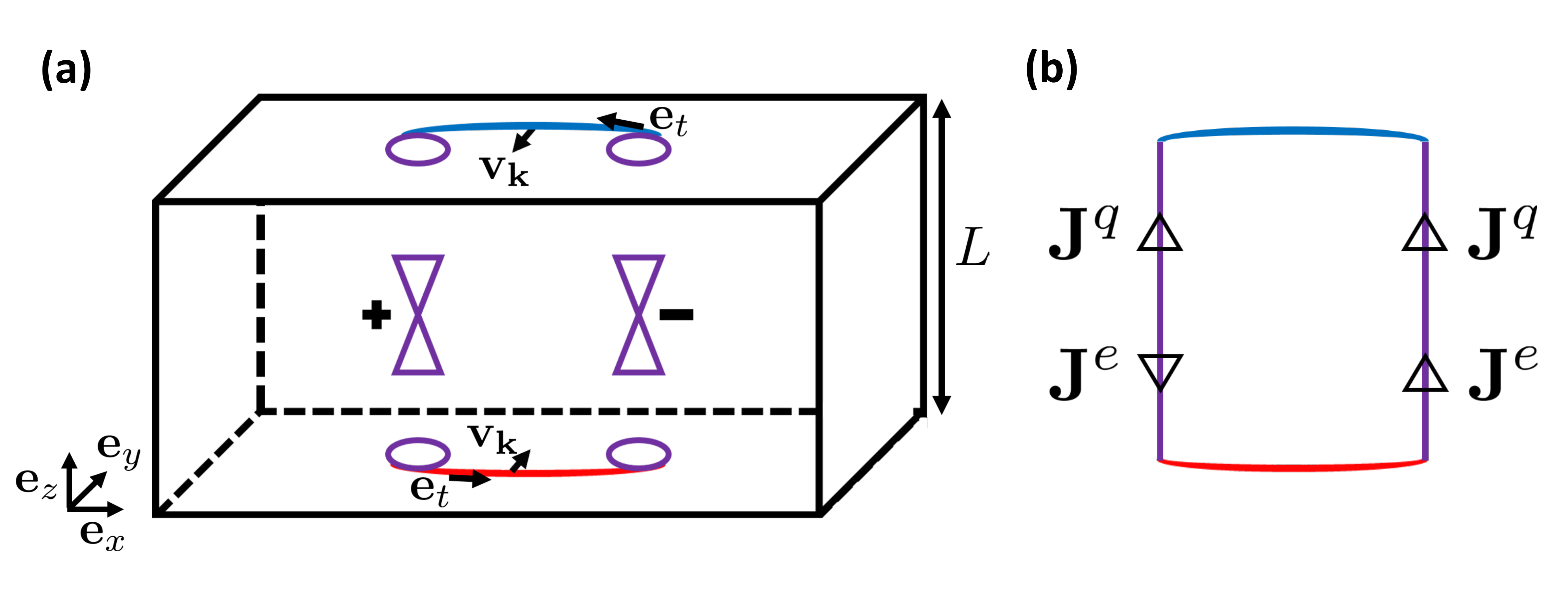}
	\caption{
	(a) Mixed real-space and momentum-space depiction of a Weyl semimetal in a slab geometry with thickness $L$ in the $z$-direction.  Bulk Weyl nodes are separated in the $x$-direction, shown in purple and labeled with their chirality $\chi = \pm 1$.  The purple ovals show that projections of the bulk Fermi pockets on the surface Brillouin zone for arbitrary chemical potential.  The Fermi arc on the top surface is shown in blue and the Fermi arc on the bottom surface is shown in red.  Unit tangent vectors $\mathbf{e}_t$ and Fermi velocities $\mathbf{v}_\bk$ are also shown for the arcs.
	(b)  Schematic of the "conveyor belt" motion of charge leading to a net heat flux.  When $\mathbf{B}$ and is aligned in the $z$-direction in (a), charge current density circulates $\mathbf{J}^{e}$ in the clockwise direction shown in a mixed real and momentum space orbit.  When $\nabla T$ is also aligned in the $z$-direction, this counter-clockwise circulation of charge leads to a net flow of heat current density $\mathbf{J}^{q}$ in the direction shown. }
	\label{slabCartoon}
\end{figure*}

In Fig. \ref{slabCartoon}a, we show a mixed real-space and momentum-space cartoon of a Weyl semimetal in a slab geometry.  We see that the bulk Weyl nodes are separated in the $x$-direction, shown in purple and labeled with their chirality $\chi = \pm 1$. The magnetic field induced flow of electrons along the arcs described by Eqn. (\ref{dndtfa2}) will lead to a transfer of electrons from right to left along the blue Fermi arc on the top surface in Fig. \ref{slabCartoon} and similarly from left to right along the red Fermi arc along the bottom surface.  In the absence of electric fields or temperature gradients, the only way the system can maintain this steady-state circulation of electrons is from bulk transport through the Fermi pockets upon which the Fermi arcs terminate.  The projections of these Fermi pockets in the surface Brillouin zone are shown in purple in Fig. \ref{slabCartoon}a.  Thus, by the continuity equation, through each bulk pocket surrounding a node with chirality $\chi$, we must have a real-space current density $\mathbf{J}_B^{\chi}$ given by
\begin{equation}
\label{sscurdens}
\mathbf{J_B^\chi} = \chi \dfrac{ e}{A} \dfrac{dN}{dt}\mathbf{e}_z,
\end{equation}
where $\frac{dN}{dt}$ is given by Eqn. (\ref{dndtfa2}).  These circulating currents are the low-field analog of those explored by the authors of Ref.\cite{pkv}.

Since the Weyl nodes come in pairs of opposite chirality, that these steady state circulating currents do not cause any net current flow along the $z$-direction in the absence of external potentials.  However, in the presence of an applied temperature gradient, entropy contributions from these circulating currents cause additive contributions to the heat current.  To see this, we consider an applied temperature gradient $\nabla T = \frac{dT}{dz}\mathbf{e}_z$ for the slab depicted in Fig. \ref{slabCartoon}, which we take to have a thickness $L$ in the $z$-direction.  Now, electrons on the top surface $z = \frac{L}{2}$ have a thermal energy density $u\left(\frac{L}{2}\right)$ and those on the bottom at $z =- \frac{L}{2}$ have thermal energy density $u\left(-\frac{L}{2}\right)$.  For a single pair of nodes, in the presence of a magnetic field $\mathbf{B} = B\mathbf{e}_z$, we have shown above that a bulk number current of electrons given by Eqn. (\ref{dndtfa2}) circulates through the bulk, driven by the Lorentz force on the top and bottom Fermi arcs.  

The flow of electrons from the Fermi arcs is topological in nature and so we only assume that scattering occurs at the surfaces of the sample.  The heat current density for $N_p$ pairs of Weyl nodes at some point $z$ in the bulk is given by
\begin{equation}
\label{fajqkzz}
J_z^{Q} = N_p L \dfrac{dN}{dt} 
\left(
u\left(z+\dfrac{L}{2}\right) - 
u\left(z-\dfrac{L}{2}\right)
\right).
\end{equation}
We expand the thermal energies $u(z\pm\frac{L}{2})$ to obtain
\begin{equation}
\label{fajqkzz2}
J_z^{Q} = \dfrac{N_p L^2}{2}\dfrac{dN}{dt}\dfrac{du}{dT}\dfrac{dT}{dz},
\end{equation}
where $\frac{du}{dT}$ is the specific heat of the bulk Weyl pockets that these thermally excited electrons from the surfaces must traverse.  We can calculate the energy density to find
\begin{equation}
\label{utot}
u = N_p \int d\E\dfrac{\E\ g(\E)}
{1+e^{\beta(\E-\mu(T))}} = \dfrac{N_p}{\pi^2(\hbar v_F)^3} \int d\E \dfrac{\E^3}{1+e^{\beta(\E-\mu(T))}}.
\end{equation}
We first consider the case of $T \gtrsim T_W$ where $\mu(T) \rightarrow 0$.  In this case, we have that 
\begin{equation}
\label{hightutot}
u \approx \dfrac{N_p (k_B T)^4}{\pi^2(\hbar v_F)^3} \int dx \dfrac{x^3}{1+e^{x}},
\end{equation}
where $\int dx \dfrac{x}{1+e^{x}} \equiv c_0 $ is a constant of order 1.  Therefore we obtain that
\begin{equation}
\label{hightdudt}
\dfrac{du}{dT} = c_0 \dfrac{N_p k_B^4 T^3}{\pi^2(\hbar v_F)^3},
\end{equation}
and so for $T \gtrsim T_W$, we obtain that 
\begin{equation}
\label{hightjq}
J_z^{Q} = \dfrac{2c_0 N_p}{\pi^2}\dfrac{e k_B^4}{\hbar^4 c} \dfrac{L^2 A k_0 T^3 B}{v_F^2}\dfrac{dT}{dz}.
\end{equation}
From Eqn. (\ref{hightjq}), we obtain the contribution of the Fermi arcs to the thermal conductivity and find it to be
\begin{equation}
\label{hightkzzz}
\kappa^{SC}_{zzz} = \dfrac{2c_0 N_p}{\pi^2}\dfrac{e k_B^4}{\hbar^4 c} \dfrac{L^2 A k_0 }{v_F^2} T^3 B,
\end{equation}
for temperatures $T \gtrsim T_W$.

\begin{figure}
	\centering
	\includegraphics[width=0.5\textwidth]
	{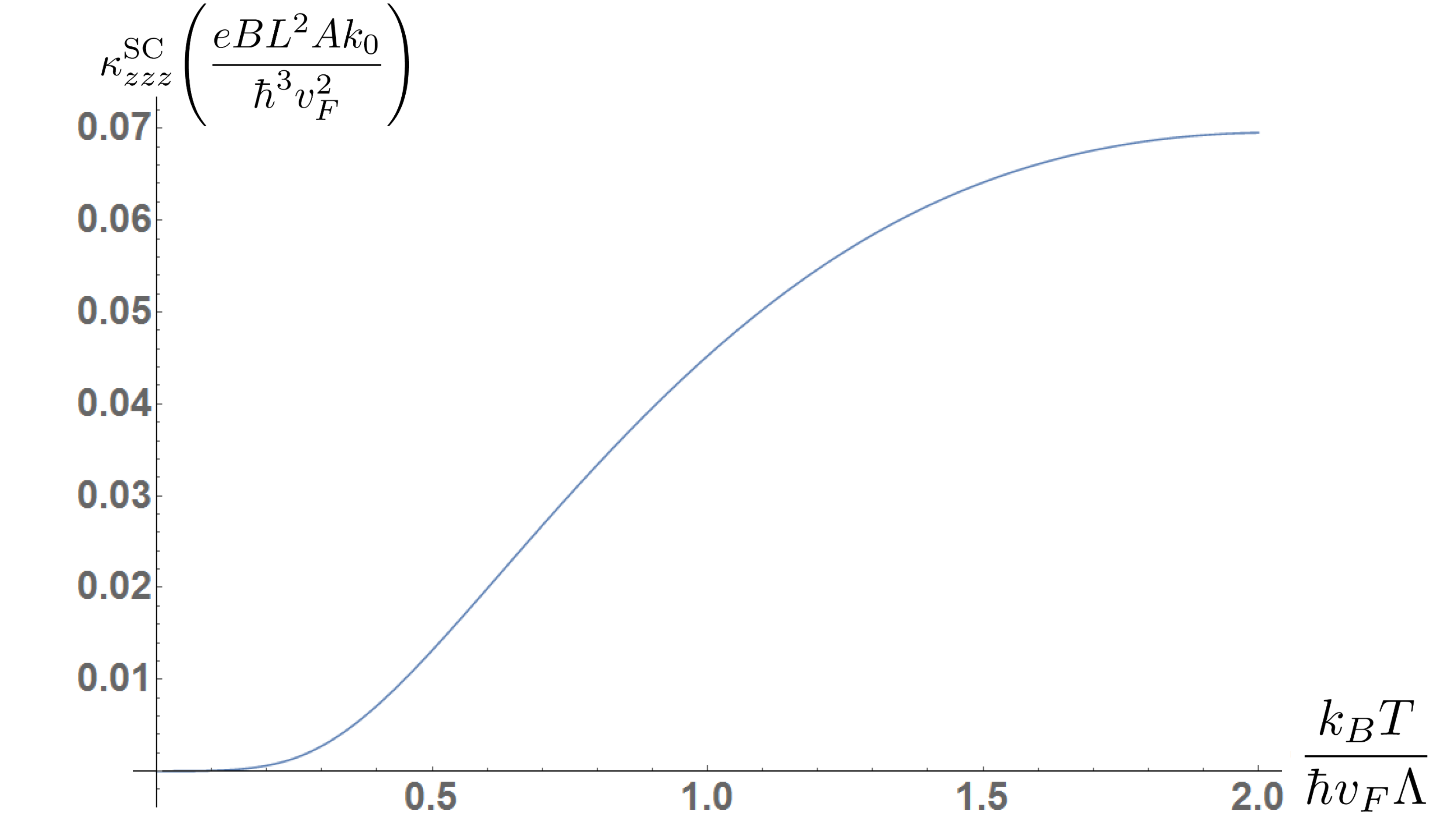}
	\caption{Fermi arc-mediated magnetothermal conductivity in the semiclassical limit as a function of temperature.  We see that at low temperatures $\kappa_{zzz}^{\textrm{SC}} \sim T^3$.  At higher temperatures, saturation of $\kappa_{zzz}^{\textrm{SC}}$ occurs as more and more bulk degrees of freedom take part in the transport of heat. At high enough temperatures, all available bulk states become involved and the magnetothermal conductivity saturates.}
	\label{scLimitKappa}
\end{figure}

In Fig. (\ref{scLimitKappa}), we show the thermal conductivity in the semiclassical limit as a function of temperature.  We see the cubic dependence at low temperature saturates as the temperature is raised.  This saturation occurs since the bulk degrees of freedom taking part in the transport of heat increases with temperature. Eventually, all available bulk states become involved and the magnetothermal conductivity saturates.
We constrast our results here with classical Drude theory for a metal where $\kappa \sim T$ and is independent of magnetic field.

\subsection{Ultra-quantum limit}

\begin{figure}
	\centering
	\includegraphics[width=0.4\textwidth]
	{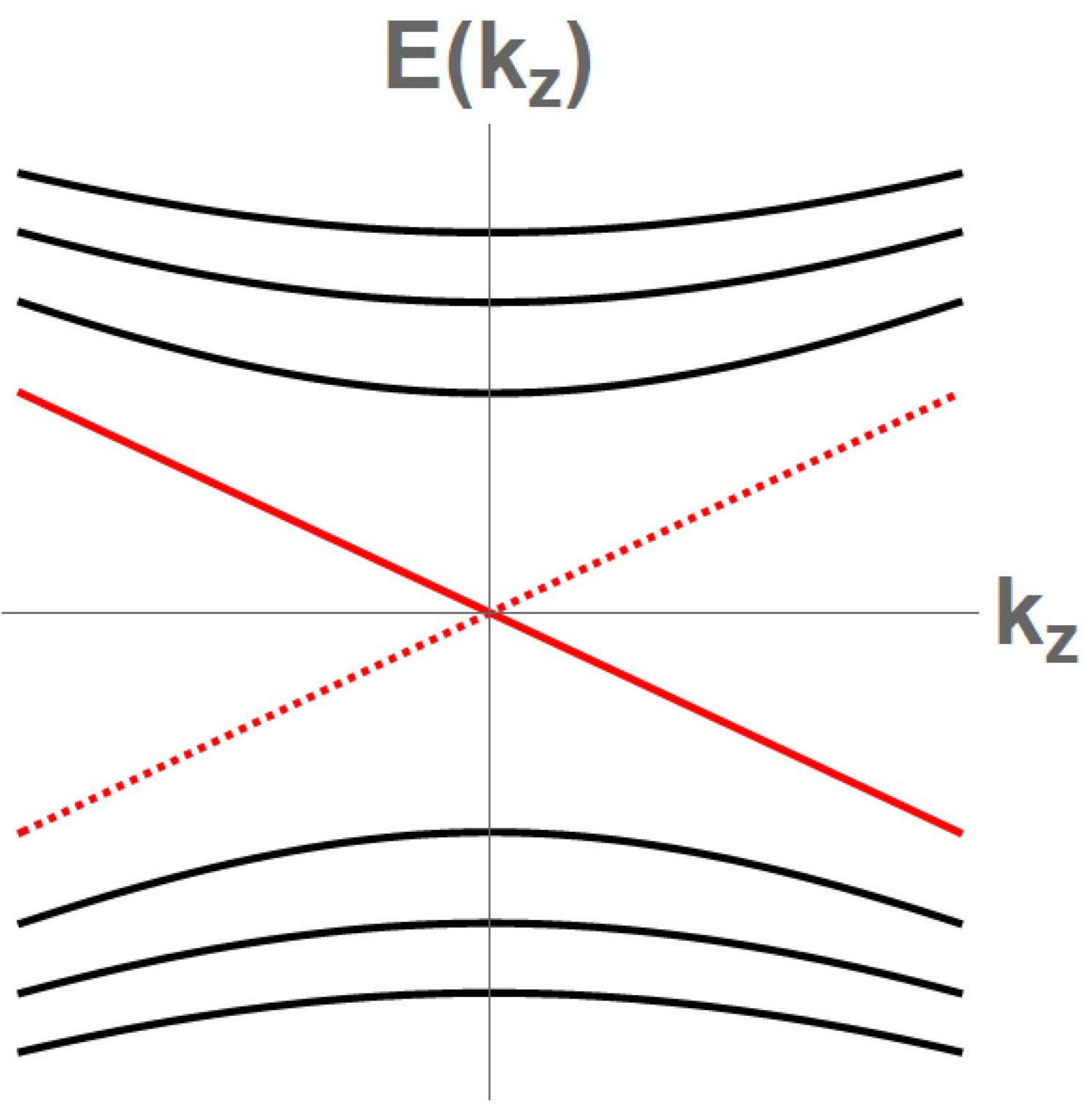}
	\caption{Schematic of the Landau levels.  Chiral $n = 0$ Landau levels are shown in red. The sign of the slope of the $n=0$ Landau levels is positive (dashed line) for $\chi = 1$ nodes and negative (solid line) for $\chi = -1$.  Non-chiral Landau levels ($n \neq 0$) are shown in black.
	}
	\label{chiralLLs}
\end{figure}

Next we analyze the limit of large magnetic fields where the energies of the Weyl nodes split into discrete Landau levels.  Due to the linear dispersion of the Weyl nodes, the Landau levels are not evenly spaced and because of their three dimensional nature, they disperse in the direction of the field in the following way
\begin{equation}
\E_{n}(k_z) = \pm \hbar v_{F} \sqrt{2|n|l_B^{-2} + k_z^2},
\label{ellnn0}
\end{equation}
for any integer $n \neq 0$ and 
\begin{equation}
\E_{0}(k_z) = \chi \hbar v_{F} k_z,
\label{elln0}
\end{equation}
for $n = 0$.  Here we have aligned the magnetic field again along the $z$-direction  such that $\mathbf{B} = B \mathbf{e}_z$.  The magnetic length is given by $l_B = \sqrt{\frac{\hbar c}{eB}}$ and $v_F$ again denotes the magnitude of the Fermi velocity.  We show a schematic of this dispersion in Fig. (\ref{chiralLLs}) where the chiral Landau levels are shown in red.  The slope of the $n = 0$ Landau level depends on the chirality of the Weyl node.

As in the semiclassical case above, on a given Fermi arc, an electron will follow the semiclassical motion of Eqn. (\ref{arckdot}).  From Ref. \cite{pkv}, we know that, in the high field limit with Landau levels in the bulk, electrons will engage in mixed momentum and real space orbits which traverse the bulk, parallel to the direction of the magnetic field.  As an electron moves along the arc, we know that it will terminate on the projection of a Weyl node which, in the presence of a large magnetic field, will be comprised of the Landau levels in Eqns. (\ref{ellnn0}-\ref{elln0}). For $k_B T \ll \frac{\hbar v_F}{l_B}$, the only state available is the zeroth Landau level given by Eqn. (\ref{elln0}).  Because of the chiral nature of these states, electrons from the arc will traverse the bulk without dissipation and emerge on the other surface.  After it reaches the other side, it will move along another Fermi arc until it reaches the opposite chirality bulk Weyl node.  It will then traverse the bulk in the opposite direction and complete the loop.

We again consider a thermal gradient in the same direction as the magnetic field such that $\nabla T = \frac{\partial T}{\partial z}\mathbf{e}_z$.  The conveyor belt motion of the electrons will again enhance the magnetothermal conductivity.  In this case however, unlike the semiclassical limit explored above, the $n = 0$ Landau levels will provide single quantum channels for heat transport.  Taking the thickness of the slab to again be $L$, we find that the heat current along the $z$-direction is given by 
\begin{equation}
\label{heatCurrqlim}
J^{q}_{z} =N_p L \dfrac{dN}{dt} 
\left(
u_{0}\left(z+\dfrac{L}{2} \right) - u_{0}\left(z - \dfrac{L}{2}\right)
\right)
=\dfrac{1}{2}N_p L^2 \dfrac{dN}{dt} \dfrac{du_{0}}{dT} \dfrac{dT}{dz},
\end{equation}
where $u_0$ is the energy density of the $n = 0$ Landau levels and $\frac{du_{0}}{dT}$ is the specific heat of a pair opposite chirality $n = 0$ Landau levels.  Again, we have assumed that continuity of charge leads to a particle current $\frac{dN}{dt}$ defined by Eqn. (\ref{dndtfa2}) through the bulk.  However, energy current is only carried by the chiral $n = 0$ Landau level.

We can calculate the internal energy
\begin{equation}
\label{intnrgqlim}
u_0 = \int d\E g_{0} (\E) f(\E)\E,
\end{equation}
where the density of states of the zero energy Landau levels is given by
\begin{equation}
\label{ll0dos}
g_{0}(\E) = CB \int_{-\Lambda}^{\Lambda} \dfrac{dk_z}{2\pi}
\bigg(
\delta(\E - |\E_0| ) + \delta(\E + |\E_0| )
\bigg),
\end{equation}
with $\E_0$ give by Eqn. (\ref{elln0}).
Here $C = \frac{N_p}{\Phi_0}$ is a normalization factor where $N_p$ is the number of pairs of Weyl nodes and $\Phi_0 = \frac{hc}{e}$ is the quantum of magnetic flux.  
$\Lambda$ is a momentum regularization which is set by the separation of the Weyl nodes in momentum-space. Then we have that
\begin{equation}
\label{intnrgqlim2}
u_0 = CB\hbar v_F\int_{-\Lambda}^{\Lambda} \dfrac{dk_z}{2\pi}
\Bigg(
\dfrac{k_z}{1+e^{\beta(\hbar v_F \Lambda-\mu)}}-\dfrac{k_z}{1+e^{\beta(-\hbar v_F \Lambda-\mu)}}
\Bigg),
\end{equation}
which can be evaluated to obtain
\begin{equation}
\label{dudtfull}
\dfrac{du_0}{dT} =\dfrac{N_p B}{\pi\Phi_0}\Bigg(4\dfrac{\hbar v_F \Lambda^2}{T}
\dfrac{1}{1+e^{-\frac{\hbar v_F \Lambda}{k_B T}}} - 8k_B \Lambda\textrm{ln}\Big( 1+e^{\frac{\hbar v_F \Lambda}{k_B T}} \Big)+
\dfrac{2k_B^2 T}{\hbar v_F}
\bigg(
\dfrac{\pi^2}{3}+4
\textrm{Li}_{2}\Big(-e^{\frac{\hbar v_F \Lambda}{k_B T}} \Big)
\bigg)
\Bigg),
\end{equation}
where $\textrm{Li}_s(z)$ is 
the polylogarithm function of order $s$, defined as 
\begin{equation}
\label{polyln}
\textrm{Li}_s(z) = \sum_{k=1}^{\infty}\dfrac{z^k}{k^s},
\end{equation}
for complex $z$ such that $|z| < 1$.

In the low temperature limit $\dfrac{k_B T}{\hbar v_F \Lambda} \ll 1$, we find that the heat capacity goes like 
\begin{equation}
\label{dudtlowt}
\dfrac{du_0}{dT} \approx \dfrac{2\pi}{3} \dfrac{N_p k_B^2}{\hbar v_F \Phi_0}T B
\end{equation}
and so we find that at low temperatures, the thermal conductivity in the ultra-quantum limit goes like
\begin{equation}
\label{kappaqlowt}
\kappa^{UQ}_{zzz} \approx \dfrac{\pi}{3} \dfrac{N_p^2 L^2 A k_0 k_B^2}{\hbar \Phi_0^2}B^2 T.
\end{equation}
Unlike the semiclassical case, we see that the thermal conductivity is quadratic in field, and now has a linear temperature dependence.  This linear dependence on temperature holds for temperatures $k_B T \ll \frac{\hbar v_F}{l_B}$.  Close to that temperature, the higher Landau levels become thermally populated, allowing for them to engage in Fermi arc-mediated magnetothermal transport as well.

The result in Eqn. (\ref{kappaqlowt}) holds up to fields high enough that the inverse magnetic length approaches the scale of the momentum cutoff $l_B^{-1} \sim \Lambda$.  For magnetic fields above $B_{c} = \frac{\hbar  c \Lambda^2 }{e}$, the coupling between the nodes cannot be neglected and a magnetic breakdown of the mixed real and momentum space orbits occurs.  In experimental realizations of Weyl semimetals, a typical value for the node separation is approximately $\Lambda \approx 0.1 \AA^{-1}$.  This allows us to estimate a typical value of $B_{c} \approx 10^3\ \textrm{T}$, well outside of the range of any laboratory fields.

\subsection{Heat capacity and entropy transport of higher Landau levels}
In order to investigate the crossover from the low-field semiclassical limit to the ultraquantum limit only involving chiral Landau levels, we include quantum effects of nonchiral Landau levels.  In this intermediate field regime, the density of states used to calculate specific heat calculated in Eqn. (\ref{intnrgqlim}) must include the higher Landau levels.  This means that Eqn. (\ref{ll0dos}) for a pair of Weyl nodes in a magnetic field becomes
\begin{equation}
\label{allLLdos}
g_{LL}(\E) = CB \int_{-\Lambda}^{\Lambda} \dfrac{dk_z}{2\pi}
\bigg(
\delta(\E - |\E_0| ) + \delta(\E + |\E_0|) +2 \sum_{n} 
\big(
\delta(\E - |\E_n| )+
\delta(\E + |\E_n| )
\big)
\bigg),
\end{equation}
where $\E_n$ is given by Eqn. (\ref{ellnn0}).  Then the additional internal energy can be calculated by
\begin{equation}
\label{llallnewnrg}
\widetilde{u} = 2 CB \sum_{n} \int_{-\Lambda}^{\Lambda} \dfrac{dk_z}{2\pi} 
\Bigg(
\dfrac{\hbar v_{F} \sqrt{|n|l_B^{-2} + k_z^2}}{1+e^{\beta(\hbar v_{F} \sqrt{|n|l_B^{-2} + k_z^2}-\mu)}}
-
\dfrac{\hbar v_{F} \sqrt{|n|l_B^{-2} + k_z^2}}{1+e^{\beta(-\hbar v_{F} \sqrt{|n|l_B^{-2} + k_z^2}-\mu)}}
\Bigg),
\end{equation}
where the total energy is now given by the sum of Eqns. (\ref{intnrgqlim2}) and (\ref{llallnewnrg}) to obtain $u_\textrm{tot} = u + \widetilde{u}$.  Unlike Eqn. (\ref{intnrgqlim2}), the expression above for $\widetilde{u}$ cannot be evaluated analytically.  Instead, we evaluate Eqn. (\ref{llallnewnrg}) numerically by introducing a regularization $N_\textrm{max}$ which cuts off the sum over Landau levels.  In the calculations below, we take the system to be at charge neutrality where $\mu = 0$.

\begin{figure*}
	\centering
	\includegraphics[width=0.99\textwidth]
	{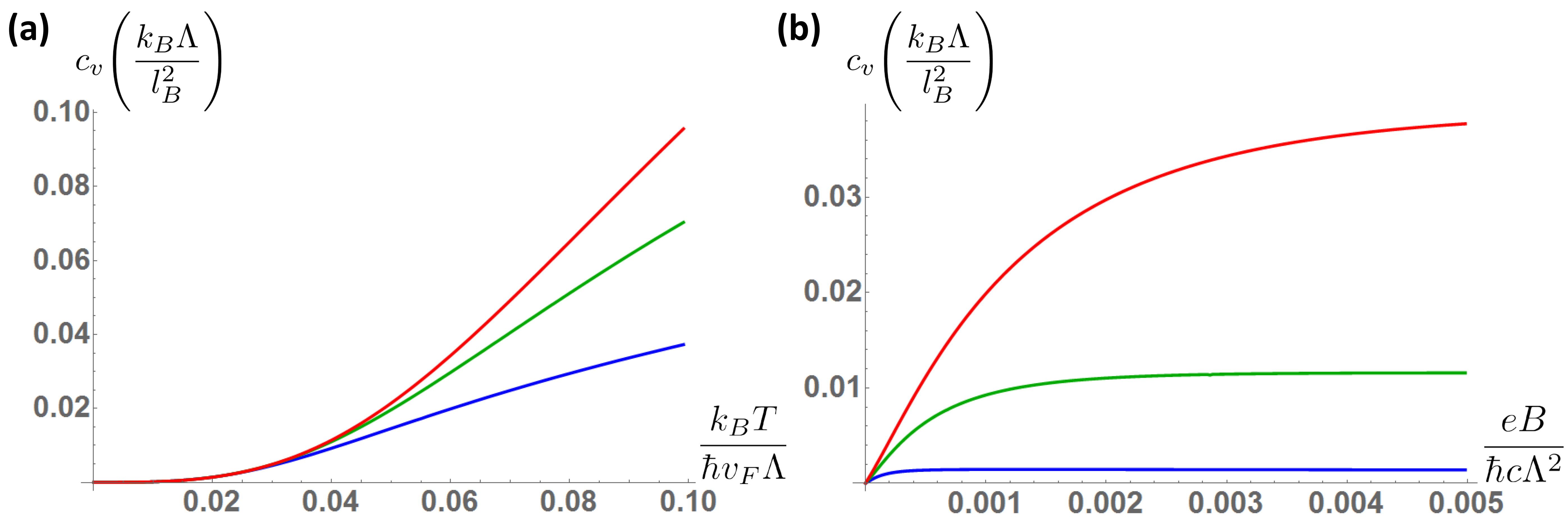}
	\caption{(a) Specific heat $c_v$ for a pair of Weyl nodes in the intermediate quantum limit versus temperature.  Calculations are performed at various fixed magnetic fields $B = 0.001 \frac{\hbar c \Lambda^2}{e}$ (blue), $B = 0.002 \frac{\hbar c \Lambda^2}{e}$ (green), and $B = 0.003 \frac{\hbar c \Lambda^2}{e}$ (red).  We have set the cutoff $N_{\textrm{max}} = 50$.
	(b) Specific heat $c_v$ for a pair of Weyl nodes in the intermediate quantum limit as a function magnetic field.  We have set the cutoff $N_{\textrm{max}} = 50$.  Here we have fixed the temperature such that $\frac{k_B T}{\hbar v_F \Lambda} = 0.02$ (blue), $\frac{k_B T}{\hbar v_F \Lambda} = 0.04$ (green), and $\frac{k_B T}{\hbar v_F \Lambda} = 0.06$ (red). 
 }
	\label{interMedQcv}
\end{figure*}

We can then calculate the heat capacity by numerically evaluating derivatives with respect to temperature. In Fig. \ref{interMedQcv}a, the total specific heat of all Landau levels $c_v = \frac{du_{\textrm{tot}}}{dT}$ is calculated as function of temperature for several fixed magnetic fields.  For low temperatures, $c_v$ increases linearly with temperature.  Around $\frac{k_B T}{\hbar v_F \Lambda} \sim 0.01$, the temperature dependence becomes cubic, before again becoming linear at high temperature.  In Fig. \ref{interMedQcv}b, $c_v$ is similarly calculated as a function of field for various fixed temperatures.  At both high and low temperatures, the field dependence of $c_v$ is linear but with different slopes.  This is precisely what we would expect from the high field ultraquantum limit in Eqn. (\ref{dudtlowt}), which is linear in field.

\begin{figure*}
	\centering
	\includegraphics[width=0.99\textwidth]
	{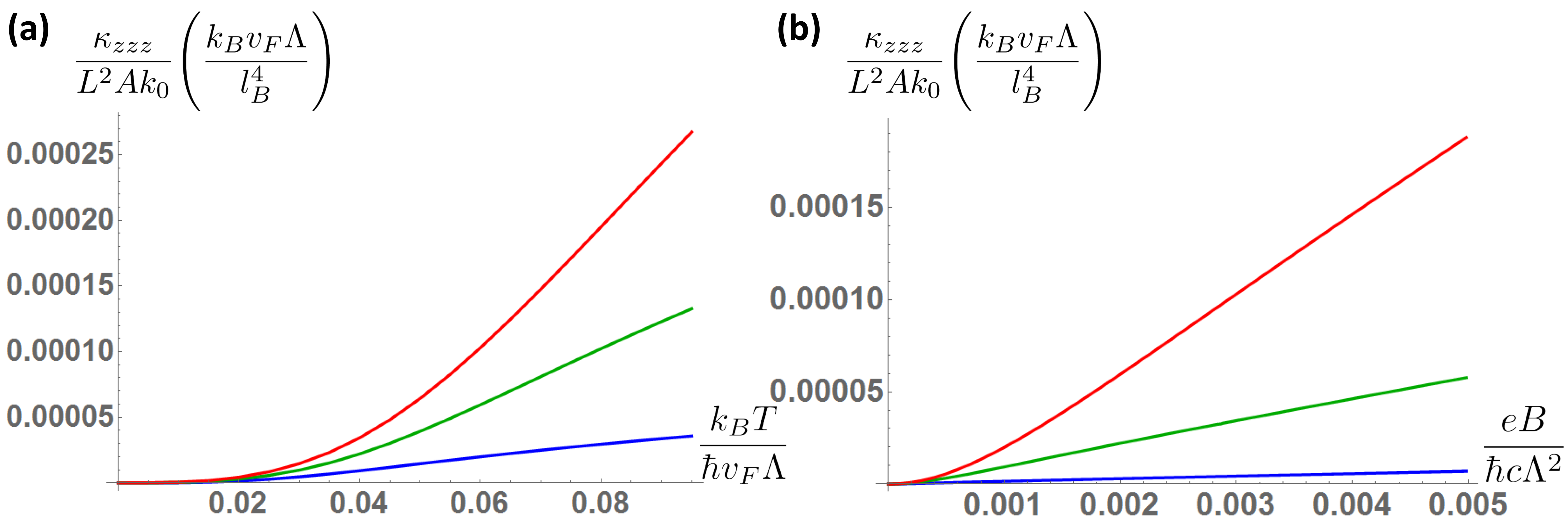}
	\caption{(a) Magnetothermal conductivity $\kappa_{zzz}$ for a single pair of Weyl nodes in the intermediate quantum limit versus temperature.  Calculations are performed at various fixed magnetic fields $B = 0.001 \frac{\hbar c \Lambda^2}{e}$ (blue), $B = 0.002 \frac{\hbar c \Lambda^2}{e}$ (green), and $B = 0.003 \frac{\hbar c \Lambda^2}{e}$ (red).  We have set the cutoff $N_{\textrm{max}} = 50$.
	(b) Magnetothermal conductivity $\kappa_{zzz}$ for a single pair of Weyl nodes in the intermediate quantum limit as a function magnetic field.  We have set the cutoff $N_{\textrm{max}} = 50$.  Here we have fixed the temperature such that $\frac{k_B T}{\hbar v_F \Lambda} = 0.02$ (blue), $\frac{k_B T}{\hbar v_F \Lambda} = 0.04$ (green), and $\frac{k_B T}{\hbar v_F \Lambda} = 0.06$ (red). 
 }
	\label{interMedQ_kappa}
\end{figure*}

After the heat capacity has been obtained, we can then calculate the arc-mediated magnetothermal conductivity.  Unlike in the case of the ultraquantum limit above, higher Landau levels do not provide the one-way channels of heat transport that the chiral $n = 0$ states do.   We now find the total heat current to be the sum of the heat current from the $n = 0$ Landau level as well as all higher Landau levels.  The total heat current for all Landau levels is given by 
\begin{equation}
\label{allLLheatcurr}
J^{q}_{z} = \dfrac{1}{2}N_pL^2 \dfrac{dN}{dt}\dfrac{du_{\textrm{tot}}}{dT}\dfrac{dT}{dz},
\end{equation}
where $c_v = \frac{du_{\textrm{tot}}}{dT}$ is the specific heat of all Landau levels as defined above.  Then the magnetothermal conductivity is given by 
\begin{equation}
\label{allLLkappazzz}
\kappa_{zzz} = \dfrac{1}{2}N_pL^2 \dfrac{dN}{dt}\dfrac{du_{\textrm{tot}}}{dT},
\end{equation}
which we evaluate numerically as above and show in Fig. \ref{interMedQ_kappa}.  In the temperature dependence of $\kappa_{zzz}$ shown in Fig. \ref{interMedQ_kappa}a, we see that the magnetothermal conductivity has a linear dependence on temperature for low and high temperatures.  In Fig. \ref{interMedQ_kappa}b, we see that the field dependence is linear for the temperature ranges shown here, where multiple Landau levels are involved in thermal transport.  This matches the calculations which we obtained semiclassically above.	

\section{Discussion and Summary}

We have seen in the presence of a magnetic field and temperature gradient, each applied perpendicular to the surface Brillouin zone of a Weyl semimetal, the Lorentz force on the Fermi arcs will lead to a conveyor belt motion of charge and a net flow of heat.  This heat flow leads to a highly anisotropic magnetothermal conductivity which has distinct behavior in the semiclassical and quantum regimes.  For relatively high temperatures and small fields, the fermi-arc mediated magnetothermal conductivity $\kappa_{zzz}$ is found to be linear in magnetic field and cubic in temperature.  On the other hand, in the ultra-quantum limit where the magnetic field is strong and temperatures are low such that only the chiral $n= 0$ Landau level is involved, we find that the Fermi arc-mediated magnetothermal conductivity is instead linear in temperature and quadratic in field.  The difference in temperature dependences can be understood by noting that in the semiclassical regime, there are many more degrees of freedom in the bulk available to thermal transport.  Once the system is the ultraquantum limit, the sole degrees of freedom available are the chiral $n =0$ Landau levels, and $\kappa_{zzz}$ is much less sensitive to changes in temperature. In the semiclassical limit, the linear magnetic field dependence comes only from the Lorentz force on the arcs.  In fields sufficiently strong enough, the quantization of the Landau levels results in  a density of states which also depends on the magnetic field.  In the ultra-quantum limit, the lone chiral $n=0$ Landau levels result in a specific heat which is quadratic in field.  As more Landau levels become populated at higher temperatures, $\kappa_{zzz}$ again becomes linear and tends toward the semiclassical limit.

In this Letter, we have focused on the so-called type I Weyl semimetals, but recently a second class, known as type II Weyl semimetals, have been theoretically predicted \cite{Soluyanov2015,Sun2015,PhysRevLett.117.056805,PhysRevB.93.201101} and observed in several candidate materials\cite{Huang2016,Wu2016,Bruno2016,Wang2016,Change1600295}.
Type II Weyl semimetals have non-vanishing density of states at the Weyl points and can be understood as the limiting case of an indirect gap semiconductor where the gap closes at the Weyl nodes.  The tilted nodes in type II Weyl semimetals lead to several distinct characteristics, perhaps most notably the lack of chiral Landau levels when the magnetic field is applied outside of the tilt cone\cite{Soluyanov2015,PhysRevLett.117.086402}.  Due to the extended nature of the Fermi pockets that meet at the Weyl nodes, lattice models\cite{mkt} are needed to accurately describe their properties rather than the continuum models used for the type I case above. 
We leave an investigation of Fermi arc mediated magnetothermal conductivity in type II Weyl semimetals for future work.

\medskip

\section{Acknowledgments} 

T. M. M. and J. P. H. were supported by the Center for Emergent Materials, an NSF MRSEC, under grant DMR-1420451.  S. J. W. was supported by the National Science Foundation Graduate Research Fellowship Program under Grant No. DGE-0822215.  N. T. acknowledges funding from NSF-DMR-1309461.

\bibliography{arcThermo_ref}

\end{document}